\documentclass[12pt]{spieman}
\usepackage{amsmath,amsfonts,amssymb}
\usepackage{graphicx}
\usepackage{setspace}
\usepackage{caption}
\usepackage{subcaption}
\usepackage{rotating}
\usepackage{hyperref}
\usepackage{hyperref}
\hypersetup{
    colorlinks=true,
    linkcolor=blue,
    filecolor=magenta,      
    urlcolor=blue,
    pdftitle={Overleaf Example},
    pdfpagemode=FullScreen,
    }
\usepackage[labelfont=bf]{caption}
\usepackage{inputenc}
\usepackage{parskip}
\usepackage{multirow}
\usepackage{indentfirst} 
\usepackage{float} 

\DeclareCaptionFormat{myformat}{\fontsize{10pt}{10pt}\selectfont#1#2#3}
\usepackage{subcaption}
\usepackage{enumitem}
\usepackage{tikz}
\usepackage{float}
\usepackage{titlesec}
\usepackage{tikz,xcolor,hyperref}
\usepackage{booktabs}
\usepackage{adjustbox}
\usepackage{amsmath}
\usepackage{algorithm}
\usepackage{algpseudocode}

\usepackage[left]{lineno}
\definecolor{lime}{HTML}{A6CE39}
\DeclareRobustCommand{\orcidicon}{
\begin{tikzpicture}
	\draw[lime, fill=lime] (0,0) 
	circle [radius=0.16] 
	node[white] {{\fontfamily{qag}\selectfont \tiny ID}};
	\draw[white, fill=white] (-0.0625,0.095) 
	circle [radius=0.007];
	\end{tikzpicture}
	\hspace{-2mm}
}
\foreach \x in {A, ..., Z}{%
	\expandafter\xdef\csname orcid\x\endcsname{\noexpand\href{https://orcid.org/\csname orcidauthor\x\endcsname}{\noexpand\orcidicon}}
}

\usepackage{tocloft}

\title {}

\cftpagenumbersoff{figure}
\cftpagenumbersoff{table}
\renewcommand{\maketitle}{
\begin{titlepage}
	\begin{center}
	\vspace{0.5cm}
	
	{\LARGE\bfseries Frequency-Guided U-Net: Leveraging Attention Filter Gates and Fast Fourier Transformation for Enhanced Medical Image Segmentation \par}
	
	\vspace{0.5cm}

	\vspace{0.5cm}
	
	\begin{tabular}{ll}
		Haytham Al Ewaidat \orcidA \textsuperscript{1,*} & Youness El Brag\orcidB \textsuperscript{2} \\
		Ahmad Wajeeh Yousef E'layan \orcidC\textsuperscript{3} & Ali Almakhadmeh \textsuperscript{4} \\
	\end{tabular}
	
	\vspace{0.5cm}
	
	\textsuperscript{1}Jordan University of Science and Technology, Faculty of Applied Medical Sciences, Department of Allied Medical Sciences-Radiologic Technology, Irbid, Jordan, 22110\\
	\textsuperscript{2}Abdelmalek Essaâdi University of Science and Technology, Faculty of Multi-Disciplinary Larache, Department of Computer Sciences, Ksar El Kebir, Morocco, 92150\\
	\textsuperscript{3,4}Jordan University of Science and Technology, Faculty of Applied Medical Sciences, Department of Allied Medical Sciences-Radiologic Technology, Irbid, Jordan, 22110
	
	\vspace{0.5cm}
 	\end{center}

	\textbf{Correspondence author:} Dr. Haytham Al Ewaidat\\
	Department of Allied Medical Sciences-Radiologic Technology\\
	Faculty of Applied Medical Sciences, Jordan University of Science and Technology\\
	PO Box 3030, Irbid 22110, Jordan\\
	Tel: (+962)27201000-26939\\
	Fax: (+962)27201087\\
	E-mail: \texttt{haewaidat@just.edu.jo}
	
	\vspace{0.5cm}
	
	\textbf{Conflict of interest:} The authors declare no conflict of interest of any type.
	
	\vspace{0.5cm}
	
	\textbf{Data availability:} The data and code for this article are available on GitHub: The code implementation can be found in our GitHub repository: \url{https://github.com/deep-matter/Attention_Filter_Gate}.
	
	\vspace{0.5cm}
	
	\textbf{Funding:} This work is supported by the Jordan University of Science and Technology, Irbid-Jordan, under grant number 20200649.
	
\end{titlepage}
}
\begin{document} 
\begin{titlepage}
	\maketitle
\end{titlepage}
\vspace*{5mm}

\begin{abstract}
\vspace{2mm}

\textbf{\fontsize{10pt}{12pt}Purpose}
Medical imaging diagnosis faces challenges, including low-resolution images due to machine artifacts and patient movement. This paper presents the Frequency-Guided U-Net (GFNet), a novel approach for medical image segmentation that addresses challenges associated with low-resolution images and inefficient feature extraction.

\textbf{\fontsize{10pt}{12pt}Approach}
In response to challenges related to computational cost and complexity in feature extraction, our approach introduces the Attention Filter Gate. Departing from traditional spatial domain learning, our model operates in the frequency domain using FFT. A strategically placed weighted learnable matrix filters feature, reducing computational costs. FFT is integrated between up-sampling and down-sampling, mitigating issues of throughput, latency, FLOP, and enhancing feature extraction.

\textbf{\fontsize{10pt}{12pt}Results}
Experimental outcomes shed light on model performance. The Attention Filter Gate, a pivotal component of GFNet, achieves competitive segmentation accuracy (Mean Dice: 0.8366, Mean IoU: 0.7962). Comparatively, the Attention Gate model surpasses others, with a Mean Dice of 0.9107 and a Mean IoU of 0.8685. The widely-used U-Net baseline demonstrates satisfactory performance (Mean Dice: 0.8680, Mean IoU: 0.8268).

\textbf{\fontsize{10pt}{12pt}Conclusion}
his work introduces GFNet as an efficient and accurate method for medical image segmentation. By leveraging the frequency domain and attention filter gates, GFNet addresses key challenges of information loss, computational cost, and feature extraction limitations. This novel approach offers potential advancements for computer-aided diagnosis and other healthcare applications.
\end{abstract}
\emph{\fontsize{8pt}{10pt}Keywords:}
Medical Segmentation, Neural Networks, U-Net Model, Attention Gate, Fast Fourier Transformation (FFT)

\fontsize{10pt}{12pt}

\section{Introduction}

Magnetic Resonance Imaging (MRI) stands as the standard technique for precisely delineating cardiac structures and etiology, guiding diagnostic and therapeutic decisions, owing to its high picture quality, excellent soft-tissue contrast, and absence of ionizing radiation \cite{ref1}. Among these structures, the left atrium (LA) holds critical importance, necessitating precise segmentation from medical images for various therapeutic applications, including cardiac illness diagnosis and therapy planning. The LA cavity, confined by a thin atrial wall with a complex structure, poses segmentation challenges \cite{ref2}. Additionally, anatomical components surrounding the atria exhibit comparable intensities, potentially confusing some segmentation algorithms \cite{ref3}. 

\begin{figure}[h]
\centering
\includegraphics[width=0.8\textwidth]{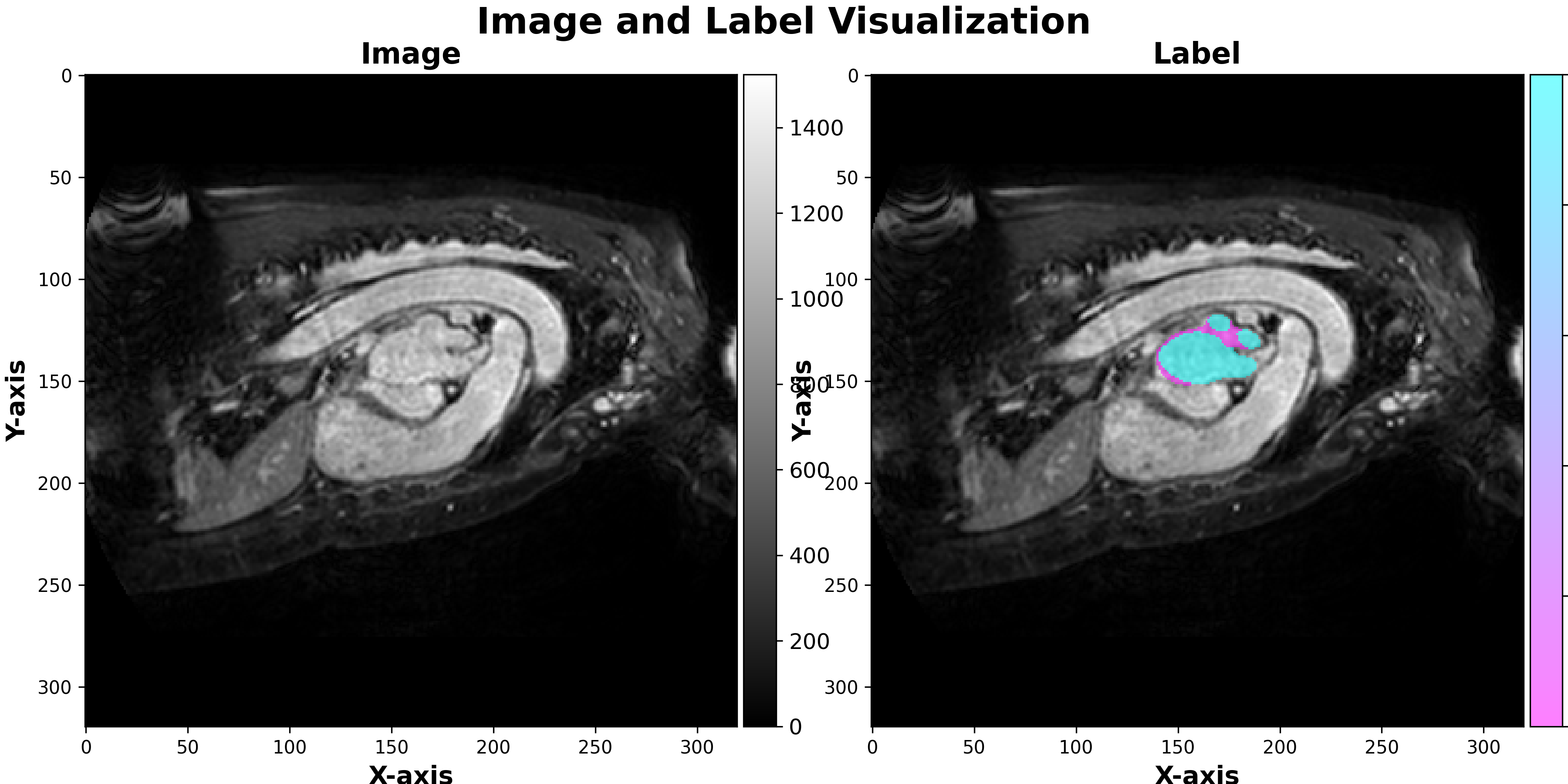}
\vspace{0.5cm}
\caption{An illustration of manual left atrium (LA) segmentation from the dataset.}
\label{fig:LA}
\end{figure}

Analyzing atrial structures, quantifying fibrosis distribution, and overcoming the challenges of manual atrial segmentation are critical aspects. Manual segmentation, a time-consuming and error-prone method, hinders accurate diagnosis and assessment. LA segmentation, measuring atrial size and function, serves as a crucial imaging marker for cardiovascular disorders such as atrial fibrillation, stroke, and diastolic dysfunction. Currently, LA segmentation is a manual and observer-dependent process, as depicted in Figure \ref{fig:LA}.

Image segmentation, dividing an image into multiple segments or regions corresponding to different objects or parts, involves various techniques and algorithms \cite{ref5}. Traditional methods, relying on thresholding or clustering, often struggle with complex or noisy images \cite{ref6,ref16}.

Recent years have witnessed the emergence of advanced image segmentation algorithms, leveraging deep learning and computer vision techniques. Convolutional Neural Networks (CNNs) trained on large datasets demonstrate the ability to recognize objects and features in images, enabling segmentation based on learned knowledge \cite{ref8}. Superpixel segmentation, another innovative method, groups similar pixels into larger regions, enhancing efficiency by reducing the number of analyzed pixels \cite{ref8,ref15}.

Despite the effectiveness of these advanced techniques, adapting them to medical imaging tasks, where training datasets are limited, remains a challenge. This challenge is especially pertinent in the context of LA segmentation in cardiac MRI.

LA segmentation is vital for assessing the risk of atrial fibrillation and planning ablation therapy \cite{ref9,ref17}. While methods like the Kalman filter show promise in handling image noise and improving segmentation \cite{ref10}, challenges persist, especially with varying LA shapes, sizes, and poor image quality post-ablation therapy \cite{ref10,ref12}.

In image segmentation, an attention filter enhances accuracy by selectively focusing on specific image regions \cite{ref13,ref18}. This technique assigns weights to input image pixels based on their importance in the segmentation task. The weight assignment, learned by a neural network during training, prioritizes pixels crucial for accurate segmentation \cite{ref13}. Applied at various stages of the segmentation pipeline, the attention filter improves overall accuracy and filters out unwanted artifacts \cite{ref18}.

Medical imaging faces challenges in disease diagnosis and analysis, aggravated by low-resolution images due to machine artifacts and patient movement. Image segmentation, identifying and localizing diseases or regions of interest within the tissue, is a critical task. Traditional segmentation methods fall short of achieving optimal results.

To overcome these limitations, artificial intelligence techniques, particularly computer vision, have been employed to enhance medical image segmentation. The U-Net model, a significant advancement in the field, has demonstrated success. However, computational cost issues during training persist.

This paper introduces a novel approach, the Frequency-Guided U-Net (GFNet), within the U-Net model for medical image segmentation. Specifically,

\section{Related Work}

Biomedical 2D image segmentation, leveraging Convolutional Neural Networks (CNNs), has demonstrated remarkable accuracy, rivaling human performance \cite{ref22}. Expanding on this success, researchers have explored the application of 3D CNNs for biomedical segmentation tasks \cite{ref19,ref20,ref21}.

The original U-Net architecture, designed for 2D image segmentation, and analogous Dense Neural Network architectures have been developed for full heart segmentation \cite{ref22}. Introducing the 3D U-Net by Cicek et al. extended U-Net to 3D segmentation \cite{ref20}. Two approaches emerged for processing volumetric data: using three perpendicular 2D slices as input and combining multi-view information for 3D segmentation \cite{ref23}, or directly replacing 2D operations in the original U-Net with their 3D counterparts \cite{ref22}. The 3D U-Net has found applications in multi-class whole heart segmentation using MR and CT data \cite{ref19}. However, GPU memory limitations have led to downsampling or predicting subvolumes of the data \cite{ref24,ref25,ref26}.

Some methods adopt a two-stage approach, extracting regions of interest (ROI) using a localization network before applying the segmentation U-Net \cite{ref27}. Our work focuses on utilizing the 3D U-Net architecture to segment 3D volumes of the whole heart obtained from 4D flow MRI, previously segmented using multi-atlas-based methods \cite{ref28,ref27,ref19}.

Automated medical image segmentation, driven by the labor-intensive and error-prone nature of manual labeling, has garnered attention in the image analysis community \cite{ref30,ref31}. CNNs, especially U-Net, exhibit promising results in tasks like cardiac MR segmentation \cite{ref30} and cancerous lung nodule detection \cite{ref24}, approaching near-radiologist level performance.

However, challenges arise when target organs exhibit significant inter-patient variation in shape and size. Fully convolutional networks (FCNs) and U-Net often resort to multi-stage cascaded CNNs \cite{ref19,ref24,ref14,ref23,ref26,ref27}. This cascade approach extracts ROIs and performs dense predictions on selected regions, leading to redundant resource usage and model parameters \cite{ref31,ref23,ref26,ref27}. Addressing this, attention gates (AGs) have been proposed as a simple yet effective solution \cite{ref19,ref23,ref26,ref27}. CNN models equipped with AGs can be trained from scratch, akin to FCN models, with AGs learning to focus on target regions automatically.

In our method, we propose a novel mechanism called the Filter Attention Gate, integrated within the U-Net model, for medical image segmentation. Our primary objective is to address the computational cost associated with model training, improve feature extraction, and overcome challenges posed by matrix multiplication in convolutional neural networks. By achieving these goals, our research aims to provide a computationally efficient algorithm for medical image segmentation, ultimately enhancing the accuracy of disease detection and localization in medical imaging.

\subsection{Semantic Segmentation}

Semantic segmentation involves splitting a digital image into multiple segments, with each segment representing a distinct object. Unlike instance segmentation, semantic segmentation does not distinguish between multiple objects of the same class. This section focuses on semantic segmentation and discusses various methods, starting with classical approaches and moving to deep learning models.

\subsection{Classical Methods for Semantic Segmentation}

Classical methods for semantic segmentation encompass a range of techniques that paved the way for contemporary approaches. These traditional methodologies, though succeeded in certain scenarios, have limitations in handling the complexity and variability of real-world images. In this section, we present an overview of classical methods commonly employed for semantic segmentation before delving into specific techniques Following : 

\begin{enumerate}
    \item \textbf{Manual Thresholding:}
    This method relies on a threshold value to create a binary image, where each pixel is either zero or one. While effective, it requires expert interaction to define thresholds and lacks generality for multiple classes.

    \item \textbf{Clustering Methods:}
    Clustering, particularly using the K-means algorithm, groups similar pixels into clusters. However, it lacks generality and requires defining variable k for each image.

    \item \textbf{Histogram-Based Methods:}
    Histogram-based methods build histograms using all pixels in the image to distinguish objects. They may struggle when peaks or valleys are absent or in a small range.

    \item \textbf{Edge Detection:}
    Edge detection identifies edges in an image based on heuristics, forming the basis for segmentation. However, it depends on heuristics and may not be viable for general cases.
\end{enumerate}

\section{Background}

The U-Net paradigm, originally introduced by Ronneberger et al. \cite{ref14}, has emerged as a pivotal convolutional neural network architecture extensively utilized in diverse image segmentation tasks. Widely embraced in medical image analysis, the U-Net addresses the complexities of training deep networks with limited annotated samples. Comprising a contracting and symmetric expanding path, it facilitates precise localization, consistently outperforming its predecessors.

\subsection{Two-Dimensional U-Net}
Initially proposed as a 2D architecture \cite{ref14}, the U-Net's adaptability to volumetric segmentation was extended by Cicek et al.\cite{ref49}. This modification replaced 2D operations with their 3D counterparts, accommodating the abundance of volumetric data in the biomedical domain. Overcoming the inefficiency of annotating large volumes slice by slice, this approach utilizes 3D convolutions, max-pooling, and upsampling layers. Notably, it mitigates challenges posed by high-resolution and large 3D data, addressing issues like resolution loss and class imbalance.

\begin{figure}[H]
\centering
\includegraphics[width=0.8\textwidth]{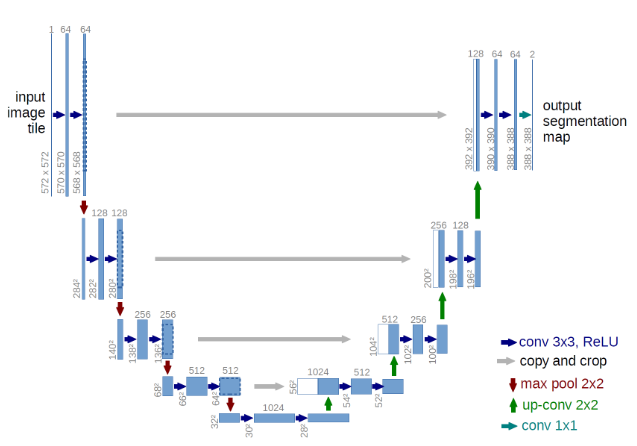}
\vspace{0.5cm}
\caption{U-Net architecture (illustrative for 32x32 pixels at lowest resolution). The blue
boxes represent a multi-channel feature map with the number of channels present on top of
each box. The size of the xy dimensions is at the bottom left of the box’s edge. Copied
feature maps are in white boxes. Each arrow denotes different operations}
\label{fig:unet}
\end{figure}

\subsection{Uncertainty Estimation using a Bayesian 3D U-Net}
Incorporating Bayesian principles into U-Net architecture, Labonte et al. \cite{ref48} introduced a method for uncertainty estimation in 3D segmentation tasks. While the sigmoid output of a U-Net offers a measure of uncertainty, it can be unreliable, especially for samples deviating significantly from the training distribution. Variational inference techniques are applied to enable accurate uncertainty estimation, a crucial aspect for interpretability, validation, and decision-making confidence. Traditional neural networks often lack this capability, leading to potential errors when confronted with data outside the training distribution.

\begin{figure}[H]
\centering
\includegraphics[width=0.8\textwidth]{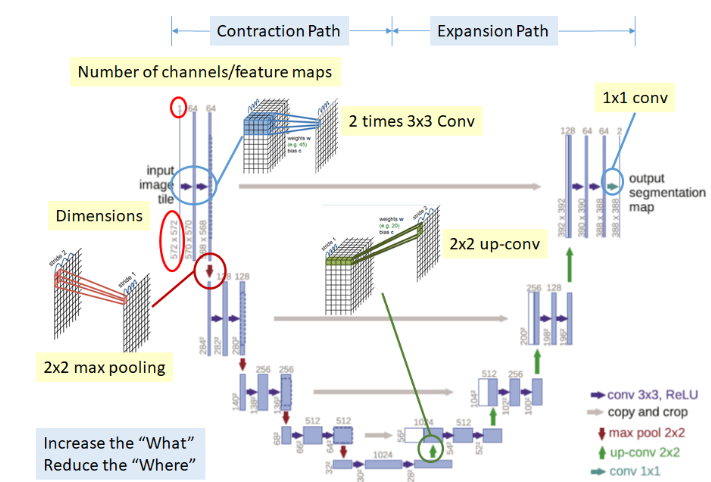}
\vspace{0.5cm}
\caption{Illustrative 3 Dimensional U-Net }
\label{fig:uner_}
\end{figure}

By embracing Bayesian 3D U-Nets, researchers have significantly advanced the segmentation framework, providing a more robust and reliable approach. This is particularly valuable in scenarios demanding precise uncertainty estimates for reliable decision-making and result interpretation.

\subsection{Attention Networks in Segmentation Tasks}

Attention networks, originally devised in Natural Language Processing (NLP), found application in handling long sentences more effectively by allowing models to focus on specific words rather than entire sentences. Initially proposed for neural machine translation \cite{ref50,ref51}, attention networks have also proven valuable in image captioning, providing interpretability by emphasizing informative features and suppressing less relevant ones, as depicted in Figure \ref{fig:cap}.

\subsubsection{Types of Attention}
Two primary attention mechanisms exist: soft attention and hard attention. Soft attention distributes attention across all regions with specific weights, while hard attention selects only one region, disregarding others. This paper primarily focuses on soft attention networks due to their differentiability and end-to-end trainability.

\subsubsection{Interpretable and Informative}
Attention networks offer transparency, revealing which regions or features the model emphasizes, providing interpretability. This section introduces attention networks, highlighting their relevance in segmentation tasks, with examples from neural machine translation and image captioning.

The subsequent sections explore various categories of attention networks used in semantic segmentation, based on how attention is generated and incorporated into the main network. The conceptual map presented here serves as a foundation for the specific attention mechanisms discussed in the following sections.

\subsection{Self-Designed Attention in Model Decoding}

\subsubsection{Attention Gate (AG)}
\begin{figure}[H]
\centering
\includegraphics[width=0.8\textwidth]{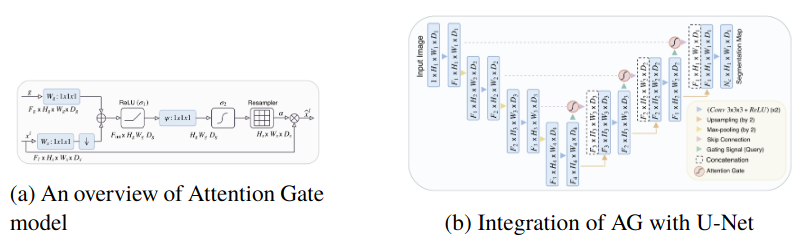}
\vspace{0.5cm}
\caption{Attention Gate model and its integration with U-Net, reproduced with permission}
\label{fig:Attention}
\end{figure}

The Attention Gate (AG) model \cite{ref18} focuses on target structures of varying shapes and sizes in medical image analysis. Integrated with CNN models like U-Net \cite{ref49}, AGs learn to suppress irrelevant regions and emphasize salient ones. The AG uses high-level contextual information to weight low-level features, enhancing computational efficiency. Figure \ref{fig:Attention}b illustrates the AG model's integration with U-Net.

To integrate AGs into U-Net, features from the decoder path weigh low-level features from the encoder. The resulting weighted features enhance the next decoding layer. Experimental evaluations on 3D CT abdominal datasets demonstrate improved segmentation performance.

\subsubsection{Pixel-wise Contextual Attention Network (PiCANet)}
\begin{figure}[H]
\centering
\includegraphics[width=0.8\textwidth]{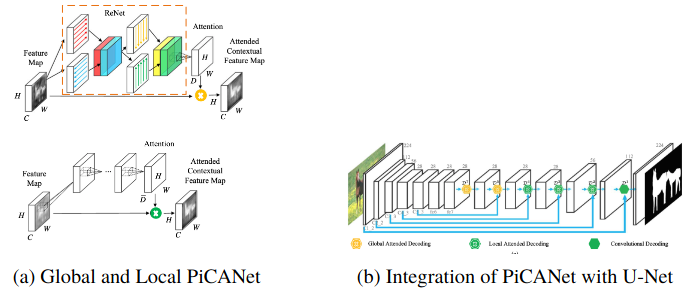}
\vspace{0.5cm}
\caption{PiCANet and its integration, reproduced with permission}
\label{fig:PiCANet}
\end{figure}

PiCANet \cite{ref54} selectively attends to informative context locations for each pixel, formulated in global and local forms. Integrated with U-Net, it allows attending to both global and local contexts. Figure \ref{fig:PiCANet}b illustrates the integration of PiCANet with U-Net.

For global PiCANet, recurrent neural networks (RNNs) capture overall features, while convolutional layers achieve local context for the local PiCANet. By incorporating PiCANet modules, U-Net effectively attends to global and local contexts, providing precise pixel-wise predictions.

\subsubsection{Discriminative Feature Network (DFN)}
\begin{figure}[H]
\centering
\includegraphics[width=0.8\textwidth]{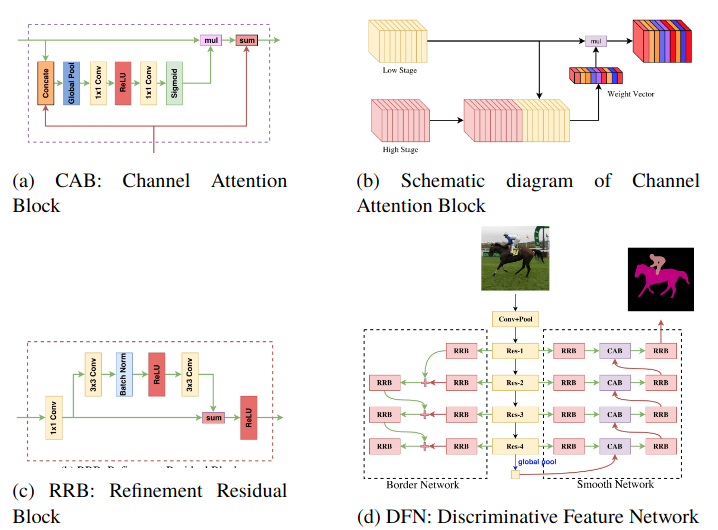}
\vspace{0.5cm}
\caption{DFN and its components \cite{ref56}, reproduced with permission.}
\label{fig:CAB}
\end{figure}

DFN \cite{ref56} addresses intra-class inconsistency and inter-class indistinction challenges in semantic segmentation. Comprising the Smooth Network and Border Network, DFN utilizes Channel Attention Blocks (CAB) and Refinement Residual Blocks (RRB), respectively. Figure \ref{fig:CAB} provides a graphical explanation of CAB and RRB.

Smooth Network captures multi-scale context, while Border Network emphasizes semantic boundaries. Experimental evaluations on PASCAL VOC 2012 and Cityscapes datasets demonstrate improved segmentation performance.

\section{Mathematical Background}
In this section, we provide a refreshed overview of the mathematical theory underpinning our exploration of complex-valued neural networks (CVNNs) and their activation functions. CVNNs, a robust modeling tool in domains where data aligns with complex numbers, present unique challenges due to analytical properties such as holomorphicity. Our focus revolves around the intricate task of formulating flexible activation functions (AFs) in the complex domain. Unlike their real counterparts, CVNNs' AFs must possess sufficient degrees of freedom to adapt their shape based on the intricacies of the training data.

While the real domain has seen extensive research in this regard, the literature on AFs for CVNNs is relatively limited. Most existing activation functions in the complex domain are developed in a split fashion, considering the real and imaginary parts separately, or rely on simple phase-amplitude techniques. To address this gap, our work introduces the Attention Filter Gate (AFG), a novel AF that combines Softmax and Sigmoid functions. The AFG is specifically tailored to handle frequencies in the Frequency Domain before executing the inverse Fast Fourier Transform (FFT). For a detailed explanation of the AFG, please refer to Section \ref{sec:method}.
\subsection{Overview of Complex-valued Neural Networks (CVNNs)}

Complex-valued neural networks (CVNNs) \cite{ref40} provide a potent modeling tool for domains where data naturally aligns with complex numbers. Despite their power, the design of CVNNs introduces challenges, notably stemming from analytical properties like holomorphicity. This study addresses the intricate task of formulating flexible activation functions (AFs) in the complex domain. We explore AFs with ample degrees of freedom to adapt their shape based on training data, an area that has received limited attention in CVNN literature. Our novel contribution is the Attention Filter Gate, a synthesis of Softmax and Sigmoid AFs tailored for handling frequencies in the Frequency Domain before the inverse FFT. Section \ref{sec:method} details this innovative approach.

\subsection{Sigmoid Activation Function for Complex Numbers}

The selection of an appropriate activation function in Equation \ref{eq:8} is notably more challenging in the complex domain due to Liouville's theorem. This theorem dictates that only constants are bounded and analytic everywhere in the complex plane. Consequently, the choice between boundedness and analyticity becomes crucial. Prior to the introduction of the sigmoid activation \cite{ref41}, early CVNN design approaches leaned towards non-analytic functions to preserve boundedness. A prevalent strategy involved applying real-valued activation functions separately to the real and imaginary parts \cite{ref42}.

\begin{equation}
g(z) = g_R(\text{Re}(z)) + i g_R(\text{Im}(z))
\label{eq:8}
\end{equation}

Here, $z$ symbolizes a generic input to the activation function in Equation \ref{eq:8}, and $g_R(\cdot)$ represents a real-valued activation function, such as the sigmoid (Equation \ref{eq:sigmoid}). This technique, termed a split activation function, is exemplified by the split-tanh activation function, showcasing magnitude and phase variations in Figure \ref{fig:acti}. Early advocates of this approach are documented in \cite{ref42} and \cite{ref43}.

Another category of frequently employed non-analytic activation functions is the phase-amplitude (PA) functions popularized by \cite{ref45, ref46}:

\begin{equation}
g(z) = \frac{z}{c + \frac{|z|}{r}}
\end{equation}

\begin{equation}
g(z) = \tanh\left(\frac{|z|}{m}\right) \exp\left\{i\varphi(z)\right\}
\end{equation}

Here, $\varphi(z)$ represents the phase of $z$, while $c$, $r$, and $m$ are positive constants typically set to 1. PA functions can be perceived as a natural extension of real-valued squashing functions, such as the sigmoid, maintaining a bounded magnitude while preserving the phase of $z$.

The Sigmoid Complex function is pivotal in processing complex inputs within our method. Conforming to the theorem of CVNNs, it computes the Sigmoid Complex activation as follows the Formalization of Sigmoid Complex based on the theorem of CVNNs he Sigmoid Complex function is formally defined as:

\begin{equation}
\text{SigmoidComplex}(x) = \begin{cases}
    \sigma(|x|) \cdot \cos(\text{angle}(x)), & \text{if } x \text{ is complex}, \\
    \sigma(x), & \text{otherwise}.
\end{cases}
\label{eq:sigmoid}
\end{equation}

In this equation, $\sigma(\cdot)$ denotes the sigmoid function, $|x|$ represents the magnitude of the complex number $x$, and $\text{angle}(x)$ signifies the angle or phase of $x$. The SigmoidComplex activation function processes a complex input $x$ by applying the sigmoid function to the magnitude while preserving the phase if $x$ is complex. For real numbers, the sigmoid function is applied directly.

The Sigmoid Complex function stands as a foundational component in our method, addressing the nuanced processing of complex-valued data within neural networks, offering a non-linear activation well-suited for complex inputs.

\subsection{Overview of Fourier Transform}
The Fourier Transform is a crucial mathematical tool that decomposes a function into its frequency components. It enables the analysis and representation of signals and functions in the frequency domain, providing valuable insights into their underlying properties. In this section, we focus on the 2D Fourier Transform.

The Fourier Theorem states that any periodic function can be expressed as a sum of sine and cosine waves with appropriate amplitudes and phases. Originating from the work of Joseph Fourier, this theorem has diverse applications in mathematics.

For a continuous function \( f(x) \) integrable over \( \mathbb{R} \), the Fourier series representation is given by:

\begin{equation*}
f(x) = \frac{a_0}{2} + \sum_{n=1}^{\infty} \left[ a_n \cos \left( \frac{2\pi n x}{T} \right) + b_n \sin \left( \frac{2\pi n x}{T} \right) \right]
\end{equation*}

Here, \( T \) is the period of the function, and the Fourier coefficients \( a_0, a_n, b_n \) are computed using the following formulas:

\begin{align*}
a_0 &= \frac{1}{T} \int_{-T/2}^{T/2} f(x) \, dx \\
a_n &= \frac{2}{T} \int_{-T/2}^{T/2} f(x) \cos \left( \frac{2\pi n x}{T} \right) \, dx \\
b_n &= \frac{2}{T} \int_{-T/2}^{T/2} f(x) \sin \left( \frac{2\pi n x}{T} \right) \, dx
\end{align*}

The inverse Fourier transform reconstructs the original function \( f(x) \) from its Fourier coefficients:

\begin{equation*}
f(x) = \sum_{n=-\infty}^{\infty} c_n e^{i 2\pi n x/T}
\end{equation*}

Here, \( c_n \) represents the Fourier coefficients, calculated using:

\begin{equation*}
c_n = \frac{1}{T} \int_{-T/2}^{T/2} f(x) e^{-i 2\pi n x/T} \, dx
\end{equation*}

The Fourier transform and its inverse are fundamental tools in signal processing, image processing, and various scientific fields. They provide insights into frequency components, facilitating tasks like noise removal, compression, and feature extraction. This discussion specifically addresses the 2D Fourier Transform applicable to two-dimensional signals, such as images.

\subsection{Discrete Fourier Transform (DFT)}
The Discrete Fourier Transform (DFT) is the discrete counterpart to the continuous Fourier Transform. It processes finite sequences of data, making it suitable for signals known only at discrete time instants. For a continuous signal \( x(t) \) sampled at intervals \( \Delta t \), the DFT of the sampled data \( x[n] \) is given by:

\begin{equation*}
X(\omega) = \sum_{n=-\infty}^{\infty} x[n] \cdot e^{-j\omega n \Delta t}
\end{equation*}

Efficient computation of the DFT is achieved using the Fast Fourier Transform (FFT) algorithm, reducing the complexity from \( O(N^2) \) to \( O(N\log N) \).

The frequency spectrum \( X(k) \) of the signal is obtained, revealing amplitudes and phases at different discrete frequencies. The inverse DFT reconstructs the original signal:

\begin{equation*}
x[n] = \frac{1}{N} \sum_{k=0}^{N-1} X(k) \cdot e^{j\omega n \Delta t}
\end{equation*}

The DFT is pivotal in digital signal processing, image processing, and scientific research, providing insights into frequency components and enabling various applications.

\subsection{Fast Fourier Transform (FFT)}
The Fast Fourier Transform (FFT) algorithm efficiently computes the Discrete Fourier Transform (DFT) of a sequence. In medical imaging, the FFT plays a crucial role in tasks such as image filtering, registration, and analysis.

For a 2D image \( f(x,y) \) with dimensions \( M \times N \), the 2D FFT \( F(u,v) \) is computed as:

\begin{equation*}
F(u,v) = \sum_{x=0}^{M-1} \sum_{y=0}^{N-1} f(x,y) e^{-i 2\pi \left(\frac{ux}{M} + \frac{vy}{N}\right)}
\end{equation*}

The inverse 2D FFT (IFFT) reconstructs the original image:

\begin{equation*}
f(x,y) = \frac{1}{MN} \sum_{u=0}^{M-1} \sum_{v=0}^{N-1} F(u,v) e^{i 2\pi \left(\frac{ux}{M} + \frac{vy}{N}\right)}
\end{equation*}

These transformations enable the analysis of spatial frequency components in medical images, contributing to tasks such as pattern identification, noise filtering, and information extraction.

The application of Fast Fourier Transform (FFT) in medical imaging is instrumental for various purposes, including image enhancement, restoration, and compression. This section provides an overview of the advantages of utilizing FFT over convolution operations in medical image processing.

\subsection{Convolution Theorem}
The Convolution Theorem establishes a fundamental connection between convolution and the Fourier Transform. For continuous functions \( f(x) \) and \( g(x) \) with Fourier Transforms \( F(\omega) \) and \( G(\omega) \) respectively, the convolution \( (f * g)(x) \) corresponds to multiplication in the frequency domain:

\begin{equation}
\mathcal{F}{f * g}(\omega) = F(\omega) \cdot G(\omega)
\end{equation}

This relationship underscores the equivalence of convolution in the time domain and multiplication in the frequency domain.

\subsection{Algorithmic Complexity Comparison}
A crucial consideration in image processing is the algorithmic complexity of operations. The FFT algorithm, with a complexity of \( \mathcal{O}(N \log N) \), outperforms convolution operations (\( \mathcal{O}(N^2) \)). This efficiency is attributed to the divide-and-conquer approach employed by the FFT, significantly reducing computational costs compared to direct convolution.

Table \ref{table:complexity} summarizes the algorithmic complexity of FFT and convolution operations, highlighting the computational advantage of FFT.

\begin{table}[H]
\centering
\begin{tabular}{p{4cm} p{5cm}}
\toprule
\textbf{Operation} & \textbf{Algorithm Complexity} \\
\midrule
FFT & \( \mathcal{O}(N \log N) \) \\
Convolution & \( \mathcal{O}(N^2) \) \\
\bottomrule
\end{tabular}
\vspace{0.5cm}
\caption{Algorithm Complexity of FFT and Convolution Operations}
\label{table:complexity}
\end{table}
\subsection{Loss Function}

In the realm of deep learning, the selection of an appropriate loss function is crucial for effective model training. For segmentation tasks, where the objective is to assign a label to each pixel or voxel in an input image or volume, specialized loss functions come into play. These functions gauge the dissimilarity between the predicted segmentation and the ground truth, steering the model towards learning precise segmentations.

One frequently employed loss function for segmentation tasks is the \textit{pixel-wise cross-entropy loss}, also recognized as \textit{binary cross-entropy loss} or \textit{softmax cross-entropy loss}. Particularly tailored for binary segmentation, this loss function computes the average cross-entropy loss across all pixels in the image.

Let's denote the predicted segmentation map as $\mathbf{P}$ and the ground truth segmentation map as $\mathbf{G}$. Both $\mathbf{P}$ and $\mathbf{G}$ are matrices of the same size as the input image, with each element representing the predicted or ground truth label of a pixel. For binary segmentation, the labels are typically encoded as 0 for background and 1 for foreground.

The pixel-wise cross-entropy loss is mathematically defined as:
\begin{equation*}
\text{CrossEntropy}(\mathbf{P}, \mathbf{G}) = - \frac{1}{N} \sum_{i=1}^{N} \left[ \mathbf{G}_i \log(\mathbf{P}_i) + (1 - \mathbf{G}_i) \log(1 - \mathbf{P}_i) \right]
\end{equation*}
where $N$ represents the total number of pixels, $\mathbf{G}_i$ is the ground truth label of the $i$-th pixel, and $\mathbf{P}_i$ is the predicted probability of the $i$-th pixel belonging to the foreground class.

This loss function penalizes the model for incorrect predictions, encouraging it to produce high probabilities for the correct class and low probabilities for the incorrect class. The logarithmic term ensures that the loss is higher when the predicted probability deviates further from the ground truth label. Minimizing this loss enables the model to generate segmentation maps closely aligned with the ground truth.

In addition to the pixel-wise cross-entropy loss, various other loss functions find common use in segmentation tasks, depending on specific requirements and problem characteristics. For instance, the \textit{categorical cross-entropy loss} is often employed in multi-class segmentation, where each pixel can be assigned one of multiple class labels.

Moreover, for tasks demanding spatial coherence and smoothness in the predicted segmentation, additional loss terms like the \textit{dice loss} or the \textit{jaccard loss} can be incorporated. These losses measure the overlap between predicted and ground truth segmentations, fostering the generation of accurate and coherent segmentations.

\subsubsection{Dice Loss}

The Dice loss stands out as another frequently utilized loss function in segmentation tasks. It gauges the overlap or similarity between predicted and ground truth segmentations, proving particularly valuable in handling imbalanced datasets where the number of background pixels far exceeds the number of foreground pixels.

The Dice loss is mathematically expressed as:
\begin{equation*}
\text{DiceLoss}(\mathbf{P}, \mathbf{G}) = 1 - \frac{2 \sum_{i=1}^{N} \mathbf{P}_i \mathbf{G}_i + \epsilon}{\sum_{i=1}^{N} \mathbf{P}_i^2 + \sum_{i=1}^{N} \mathbf{G}_i^2 + \epsilon}
\end{equation*}
where $\epsilon$ is a small constant added for numerical stability. The Dice loss ranges from 0 to 1, with 0 indicating no overlap between predicted and ground truth segmentations, and 1 indicating perfect overlap.

The Dice loss encourages the model to generate segmentations with high overlap with the ground truth, promoting accurate segmentation of both foreground and background regions. Its efficacy shines through, especially when dealing with imbalanced class distributions.

In practice, a common strategy involves combining the cross-entropy loss and the Dice loss to create a hybrid loss function. This amalgamation leverages the strengths of both loss functions, fostering accurate pixel-wise classification while also promoting spatial coherence and overlap with the ground truth.

\subsubsection{Cross Entropy Loss}

For multi-class segmentation tasks, the \textit{cross-entropy loss}, also known as the \textit{softmax cross-entropy loss}, takes center stage. This loss function is adept at scenarios where each pixel can be assigned one of multiple class labels.

Let's represent the predicted segmentation map as $\mathbf{P}$ and the ground truth segmentation map as $\mathbf{G}$. Both $\mathbf{P}$ and $\mathbf{G}$ are matrices of the same size as the input image, with each element signifying the predicted or ground truth label of a pixel. The cross-entropy loss is mathematically defined as:
\begin{equation*}
\text{CrossEntropy}(\mathbf{P}, \mathbf{G}) = - \frac{1}{N} \sum_{i=1}^{N} \sum_{c=1}^{C} \mathbf{G}_{ic} \log(\mathbf{P}_{ic})
\end{equation*}
where $N$ is the total number of pixels, $C$ is the number of classes, $\mathbf{G}_{ic}$ is the ground truth label of the $i$-th pixel for the $c$-th class, and $\mathbf{P}_{ic}$ is the predicted probability of the $i$-th pixel belonging to the $c$-th class. This loss function penalizes the model for incorrect predictions and encourages it to produce high probabilities for the correct class.

\subsubsection{BCE-DICE Loss}

In the segmentation domain, employing an effective loss function is pivotal to capturing crucial features and preserving spatial information. The Binary Cross-Entropy (BCE) loss and the Dice loss come together in the BCE-Dice loss, providing a hybrid approach that incorporates both spatial and label-wise information for accurate segmentation.

\begin{enumerate}
\item \textbf{Binary Cross-Entropy (BCE) Loss:} The BCE loss quantifies the difference between the predicted probability and the ground truth label. Widely used in binary classification tasks, it is computed by taking the negative logarithm of the predicted probability for the correct label. The formulation is given by Equation \ref{eq:3}.

\item \textbf{Dice Loss:} The Dice loss measures the overlap between the predicted segmentation and the ground truth segmentation. It is calculated by evaluating the ratio of twice the intersection of the two segmentations to the sum of the pixels in both segmentations. The Dice loss aids in evaluating the similarity between the predicted and ground truth segmentations. The mathematical expression is provided by Equation \ref{eq:4}.
\end{enumerate}

By combining the BCE loss and the Dice loss, the BCE-Dice loss allows for the incorporation of both spatial and label-wise information during the segmentation task. The BCE loss encourages accurate pixel-wise classification, while the Dice loss motivates precise

In conclusion, the Fast Fourier Transform proves to be an efficient computational approach for medical image processing, offering significant speed-ups and reduced computational costs compared to convolution operations.

\section{\fontsize{12pt}{12pt}Materials and Methods}

In this section, we present the Frequency-Guided U-Net (GFNet) architecture with the Attention Filter Gate Network (AFGN). The architecture is designed to enhance U-Net models for advanced image processing tasks. The conceptual foundation and mathematical intricacies are detailed, along with the code snippets for the key components.

\subsection{Frequency-Guided U-Net (GFNet)}

GFNet is an innovative architecture aimed at improving the performance of U-Net models through the integration of the Attention Filter Gate Network (AFGN). The key idea behind GFNet is to exploit frequency domain interactions to capture both long-term and short-term features. The architecture eliminates inductive biases while maintaining log-linear computational complexity, aligning with the prevailing trends in deep learning.

\begin{figure}[h]
\centering
\includegraphics[width=0.8\textwidth]{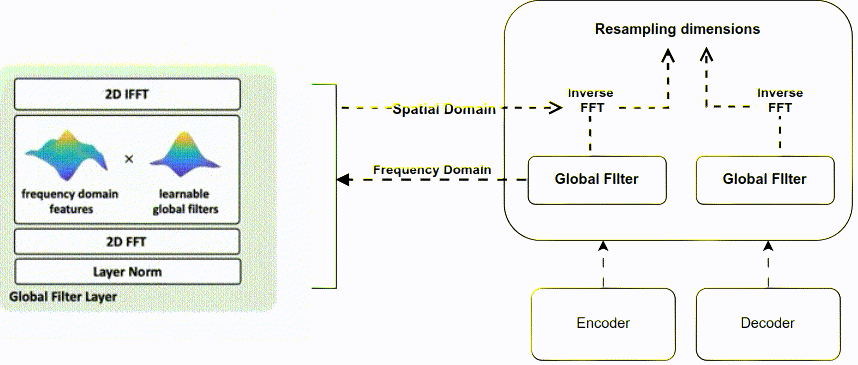}
\vspace{0.5cm}
\caption{Overview of the Attention Filter Gate Network (AFGN) layer, enhancing performance by focusing on local features based on global context in neural networks.}
\label{fig:methods}
\end{figure}

Figure \ref{fig:methods} illustrates the AFGN layer within the GFNet architecture. The AFGN layer replaces traditional attention-gate mechanisms with operations in the frequency domain. This allows for the exploration of interactions among spatial locations in the frequency domain, providing a more efficient and effective way to capture complex features.

\subsubsection{AFGN Operations}

The core operations of the Attention Filter Gate Network include the use of global filters and the integration of Fast Fourier Transform (FFT). The AFGN layer is seamlessly integrated into the U-Net model through three essential operations:

\begin{enumerate}
  \item \textbf{2D Discrete Fourier Transform (DFT):} For spatial-to-frequency conversion.
  \item \textbf{Element-wise Multiplication:} With learnable global filters.
  \item \textbf{2D Inverse Fourier Transform (iFFT):} For frequency-to-spatial mapping.
\end{enumerate}

The use of FFT significantly enhances the efficiency of the Attention Filter Gate, enabling compatibility with larger feature maps and CNN-style hierarchical architectures without requiring extensive modifications.

\subsection{Attention Filter Gate (AFG)}

To seamlessly integrate the AFGN layer into the U-Net model, the attention-gate sub-layer is replaced with the operations outlined below. This ensures the GFNet architecture's adaptability to diverse image processing tasks without substantial modifications to the existing U-Net structure.

In this section, we present a comprehensive research approach and study design underlying the GFNet architecture. We delve into the mathematical foundation, advanced concepts, and practical applications, offering a holistic view of the research methodology.

The research leverages the 2D Fourier Theorem as a foundational element, allowing the representation of periodic functions as combinations of sine and cosine waves. This theorem finds relevance in scientific and engineering domains, including signal analysis and image processing.

Expanding upon this foundation, we explore advanced concepts such as the Fast Fourier Transformation (FFT) and the Convolution Theorem. These mathematical tools play pivotal roles in tasks like image compression, denoising, and feature extraction.

Introducing two essential components: the Global Filter and the Attention Filter. These components seamlessly integrate to form the Spatial Attention Filter Gate (SAFG) approach. The Global Filter transforms input global feature maps, while the Attention Filter selectively attends to specific features within the context of global information.

The interplay of these components forms the foundation of the SAFG approach, which will be further explored in subsequent sections.

This section details the study setting, focusing on the Global Filter Layer, a critical component of the research. The Global Filter Layer operates through a series of operations, including a 2D FFT, element-wise multiplication, and a 2D iFFT. These operations collectively contribute to shaping the frequency components of the transformed image.

\begin{algorithm}
    \caption{Global Filter Layer}
    \begin{algorithmic}[1]
        \Function{GlobalFilter}{$x$}
            \State \textbf{Input:} $x$: input tensor
            \State \textbf{Output:} $x$: output tensor, $Frequency\_dom$: frequency domain tensor
            \State $W \gets (out\_channel // 2) + 1$
            \State $H \gets in\_channel$
            \State $complex\_weight \gets \text{Parameter}(torch.randn(H, W, dim, 2, dtype=torch.float32) * 0.02)$
            \State $x \gets \text{rfft2}(x, dim=(1, 2), \text{norm}='ortho')$
            \State $weight \gets \text{view\_as\_complex}(complex\_weight)$
            \State $Frequency\_dom \gets x * weight$
            \State $x \gets \text{irfft2}(Frequency\_dom, s=(H, W), dim=(1, 2), \text{norm}='ortho')$
            \State \textbf{Return} $x$, $Frequency\_dom$
        \EndFunction
    \end{algorithmic}
\end{algorithm}

The Global Filter Layer is integral to the research approach, contributing to the enhancement of spatial attention within neural network architectures.

This section focuses on the Attention Filter Gate (AFG), a pivotal component within GFNet that significantly enhances performance across various domains, including image recognition and natural language processing.

The Attention Filter Gate comprises two primary components: the gate tensor and the weight tensor. These components undergo transformations from the spatial domain to the frequency domain via Fast Fourier Transform (FFT) operations.

The frequency domain representations are obtained for the gate and weight tensors, facilitating the computation of the attention matrix through element-wise multiplication and the application of the softmax function. The attention matrix is then converted back to the spatial domain using the inverse Fast Fourier Transform (iFFT) operation.

\begin{algorithm}
\caption{Attention Filter Gate}
\begin{algorithmic}[1]
\Procedure{AttentionFilterGate}{$F_g, F_l, F_{\text{int}}, \text{dim}, g, x$}
    \State \textbf{Input}:
        \Statex $F_g$ - Frequency for gate computation
        \Statex $F_l$ - Frequency for weight computation
        \Statex $F_{\text{int}}$ - Internal frequency parameter
        \Statex $\text{dim}$ - Dimension parameter
        \Statex $g$ - Global feature map
        \Statex $x$ - Feature map
    \State \textbf{Output}:
        \Statex $out$ - Filtered output
    
    \State $x1, g1_{\text{freq}} \gets \text{GlobalFilter}(g, F_l, F_{\text{int}}, \text{dim})$
    \State $x2, x1_{\text{freq}} \gets \text{GlobalFilter}(x, F_g, F_{\text{int}}, \text{dim})$
    \State $\text{norm\_freq} \gets \text{Variance}(g1_{\text{freq}})$
    \State $\text{atten} \gets \text{SigmoidComplex}\left(\frac{g1_{\text{freq}}^T \cdot x1_{\text{freq}}}{\sqrt{2 \cdot \pi \cdot \text{norm\_freq}}}\right)$
    \State $\text{inverse\_atten} \gets \text{InverseFastFourierTransform}(\text{atten}, s=(H, W), \text{dim}=(2, 3), \text{norm}='ortho')$
    \State $out \gets \text{LayerNorm}(x + \text{inverse\_atten})$
    \State \textbf{Return} $out$
\EndProcedure

\Procedure{Variance}{$x$}
    \State \textbf{Input}:
        \Statex $x$ - Input tensor
    \State \textbf{Output}:
        \Statex $x$ - Variance
    
    \State $x \gets \text{torch.var}(x, \text{dim}=(2, 3), \text{keepdims}=True)$
    \State \textbf{Return} $x$
\EndProcedure
\end{algorithmic}
\end{algorithm}

The Attention Filter Gate operations are encapsulated in the \texttt{AttentionFilterGate} algorithm, showcasing the complexity of computations involved.

A comprehensive complexity analysis table provides an overview of the computational complexity in terms of Floating-Point Operations (FLOPs) and the number of parameters for various operations within GFNet. This analysis aids in understanding the resource requirements for the proposed architecture.

\begin{table}[H]
\centering
\begin{tabular}{p{3.5cm}p{5cm}p{5cm}}
\toprule
Operation & Complexity (FLOPs) & Parameters \\
\midrule
Convolution & $O(k^2HW D)$ & $k^2D$ \\
(AFGN) & $O(HW D^2 + H^2W^2D)$ & $2HW(F_l + F_g + F_{\text{int}}) + 4D^2$ \\
(AG) & $O(HW^2D)$ & $HW$ \\
\bottomrule
\end{tabular}
\vspace{0.5cm}
\caption{Complexity Analysis}
\end{table}

The complexity analysis provides a clear understanding of the computational demands for various operations within GFNet.

\section{Comparative Analysis: Attention Gate vs. Attention Filter Gate}

This section presents a thorough comparative analysis of two crucial techniques in image processing: Attention Gate and Attention Filter Gate. These techniques significantly contribute to improving the quality and effectiveness of image-based operations.

The Attention Filter Gate, a vital component of the Global Filter Network (GFNet) architecture, is designed to capture long-term spatial dependencies in the frequency domain. This approach stands out due to the incorporation of learnable weights as trainable parameters woven into the U-Net model's architecture and the utilization of the Fast Fourier Transform (FFT) in signal processing. The FFT accelerates the multiplication process, resulting in substantial improvements in image and audio signal quality.

\begin{figure}[H]
\centering
\includegraphics[width=0.8\textwidth]{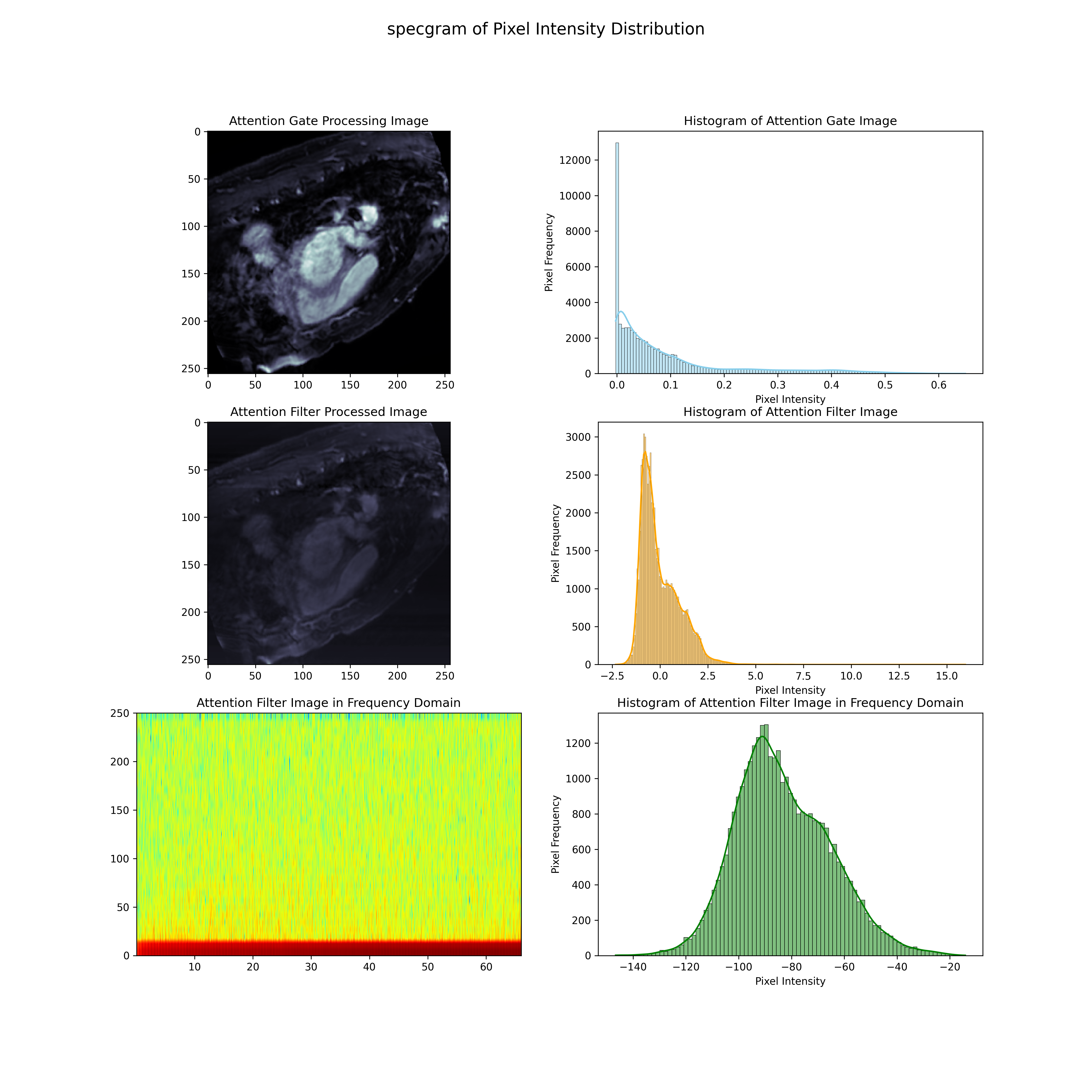}
\vspace{0.5cm}
\caption{Comparison: Attention Gate vs. Attention Filter Gate - Learning from Frequency vs. Spatial Domain based on Z-normalized Pixel Distribution in the Frequency Domain.}
\label{fig:freq}
\end{figure}

In contrast, the Attention Gate is renowned for its effectiveness in various deep learning applications. It excels in selectively focusing on specific input data regions, enhancing feature extraction and discrimination. Within the U-Net model, the Attention Gate proves invaluable for tasks like image segmentation, enabling the network to concentrate on relevant image regions while suppressing noise and irrelevant details.

To comprehensively compare these techniques, we employ a multifaceted approach, combining both qualitative and quantitative analyses.

\subsection{Qualitative Analysis: Z-normalized Histograms}

We delve into the qualitative aspects by examining the impact of these techniques on pixel intensity distributions. Z-normalization, a statistical technique, transforms pixel intensity distributions into a standard Gaussian distribution in the frequency domain. This comparison of Z-normalized histograms provides insights into their influences on image characteristics such as contrast, brightness, and overall pixel value distribution.

This multifaceted comparison offers a comprehensive understanding of how Attention Gate and Attention Filter Gate impact image processing tasks in the frequency domain, shedding light on their applicability in various domains.

\section{Data Collection and Preprocessing}

The dataset for this task is in a 3D full volume format, requiring essential preprocessing steps:

\begin{itemize}
  \item \textbf{2D Slice Extraction:} Extract 2D slices corresponding to the target label from the 3D full volume dataset.
  
  \item \textbf{Crop Region of Interest (ROI):} Crop extracted slices to retain only the region of interest (ROI), including the left atrium and pulmonary veins, crucial for focusing the model on relevant anatomical structures.
  
  \item \textbf{Image Resizing:} Resize extracted slices to a fixed dimension to ensure uniformity in input data size for efficient processing.
  
  \item \textbf{Z-normalization:} Normalize pixel values within resized slices based on mean and standard deviation to standardize the data, making it amenable to machine learning algorithms.
  
  \item \textbf{Data Augmentation:} Apply various data augmentation techniques such as rotation, scaling, and elastic transformations to diversify the dataset, enhancing the model's robustness.
\end{itemize}

These preprocessing steps are essential to prepare the data for subsequent segmentation tasks, enabling the model to focus on relevant anatomical structures, standardize pixel values, and increase dataset variability for improved performance.

\begin{figure}[H]
\centering
\includegraphics[width=0.8\textwidth]{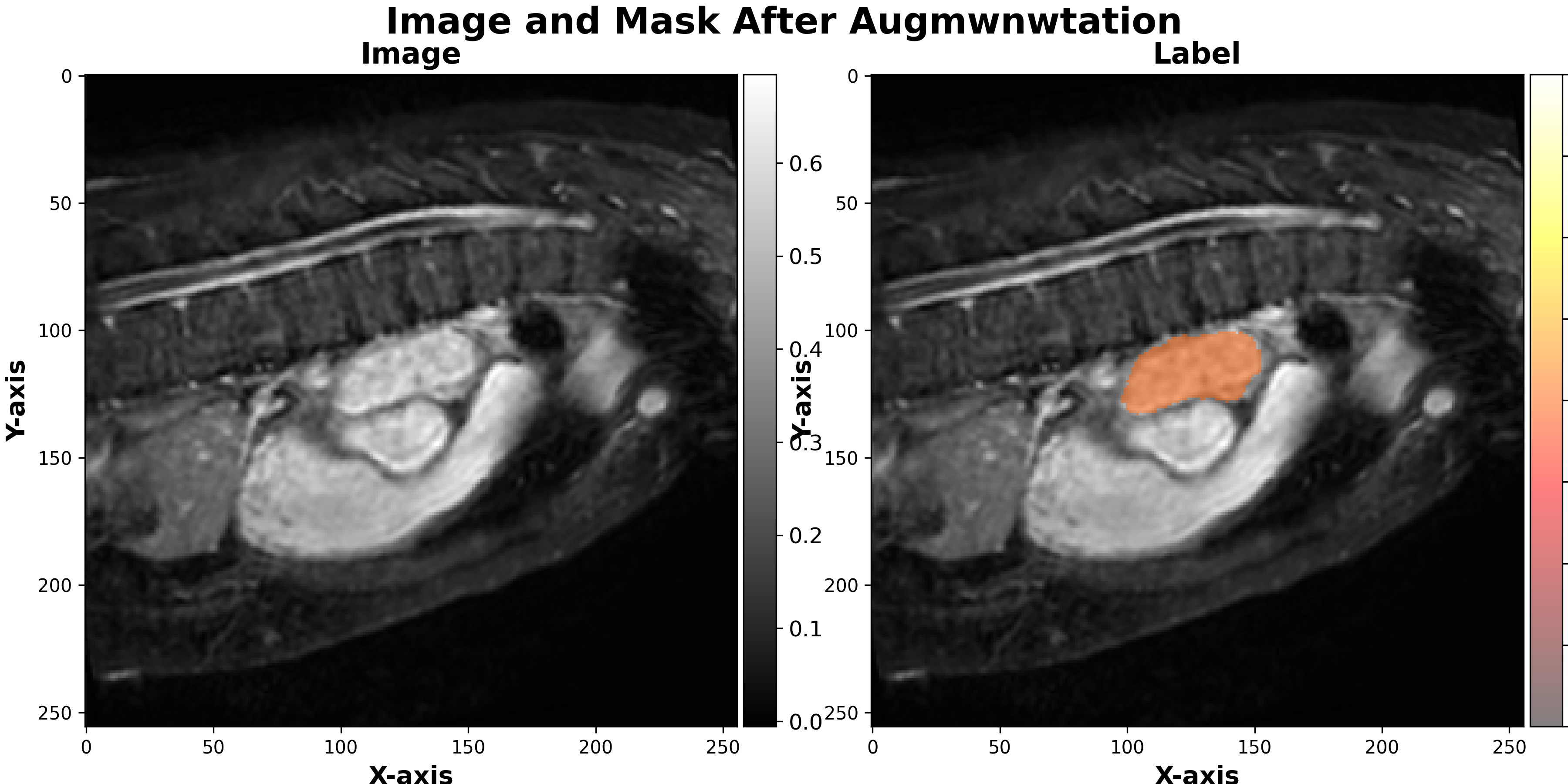}
\vspace{0.5cm}
\caption{Overlapping Labels within Processed Images}
\label{fig:fig1}
\end{figure}

\section{Experimental Setup}

This section details the experiments conducted to evaluate the model's performance, summarizing the experimental setup and parameters.

\begin{table}[H]
\centering
\caption{Experimental Setup}
\vspace{0.5cm}
\begin{tabular}{l|l}
\toprule
Parameter & Value \\
\midrule
Image Size & 256 \\
GPU & GPU T4 x2 \\
Number of Filters & 64 \\
Dropout in Encoder & False \\
Dropout in Decoder & True \\
Category & Atrium \\
Number of Channels & 1 \\
Attention Mechanism & Filter \\
Number of Hidden Layers & 1 \\
Number of Classes & 1 \\
Initializer Range & 0.02 \\
Monitor CAM & GradCAM++ \\
Weight Decay & 0.01 \\
Learning Rate (lr) & 0.001 \\
\bottomrule
\end{tabular}
\end{table}

These experiments aim to provide a comprehensive evaluation of the proposed approach for atrium segmentation, considering segmentation accuracy, computational efficiency, and the impact of different hyperparameters.

\section{Results Analysis}

This section analyzes the results of Attention Filter Gate, Attention Gate, and U-Net models using Mean Dice and Mean Intersection over Union (IoU) metrics.

\begin{table}[H]
\centering
\caption{Model Performance Comparison}
\vspace{0.5cm}
\begin{tabular}{lcc}
\toprule
Model & Mean Dice & Mean IoU \\
\midrule
Attention Filter Gate & 0.8366 & 0.7962 \\
Attention Gate & 0.9107 & 0.8685 \\
U-Net & 0.8680 & 0.8268 \\
\bottomrule
\end{tabular}
\end{table}

The results indicate that the Attention Gate model achieved the highest Mean Dice and Mean IoU scores, demonstrating superior performance in image segmentation tasks. The Attention Filter Gate model also performed well, with competitive scores, showcasing its effectiveness in capturing spatial dependencies in the frequency domain.

\begin{figure}[H]
\centering
\includegraphics[width=0.8\textwidth]{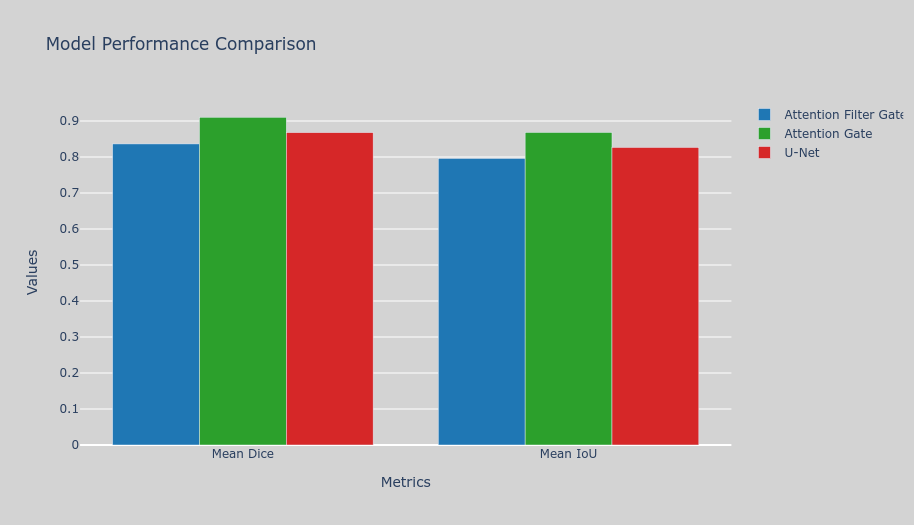}
\vspace{0.5cm}
\caption{Comparison of Dice and IoU Metrics for all Models}
\label{fig:box}
\end{figure}

To provide a more in-depth analysis, comparing metrics:

\begin{itemize}
\item \textbf{Mean Dice:} Attention Gate achieved the highest Mean Dice, indicating precise segmentation.

\item \textbf{Mean IoU:} Attention Gate exhibited the highest Mean IoU, implying better overall segmentation performance.

\end{itemize}

\begin{figure}[H]
\centering
\begin{subfigure}{0.8\textwidth}
  \centering
  \includegraphics[width=\linewidth]{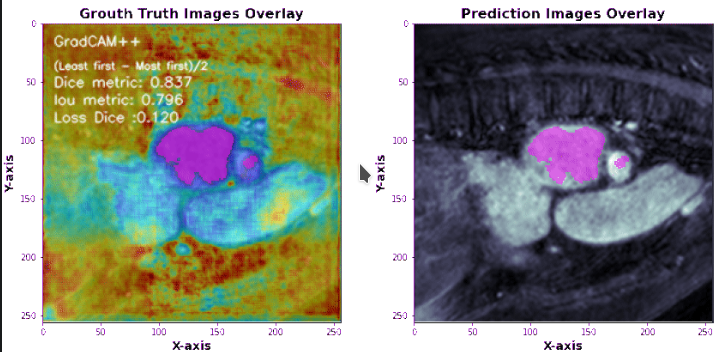}
  \vspace{0.5cm}
  \caption{Attention Filter Gate Prediction Overlap within Mask}
\end{subfigure}
\begin{subfigure}{0.8\textwidth}
  \centering
  \includegraphics[width=\linewidth]{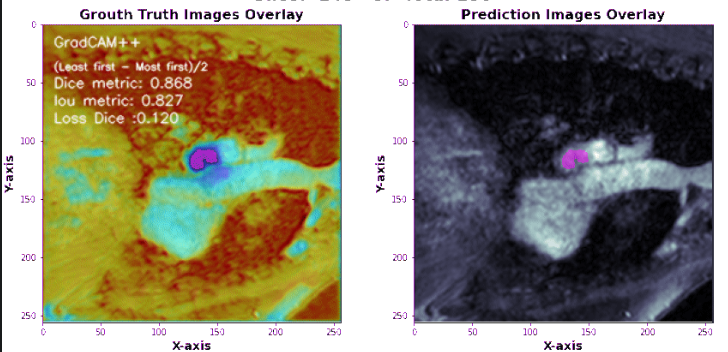}
  \vspace{0.5cm}
  \caption{U-Net Prediction Overlap within Mask}
\end{subfigure}
\vspace{0.5cm}
\caption{GradCAM++ Visualization - Attention Filter Gate vs. U-Net}
\label{fig:segmentation}
\end{figure}

Additionally, the GradCAM++ technique visualizes regions of interest influencing the model's segmentation decisions, revealing high-importance areas in input images.

\begin{figure}[H]
\centering
\begin{subfigure}{0.8\textwidth}
  \centering
  \includegraphics[width=\linewidth]{figures/Filter.png}
  \vspace{0.5cm}
  \caption{Attention Filter Gate Prediction Overlap within Mask}
\end{subfigure}
\begin{subfigure}{0.8\textwidth}
  \centering
  \includegraphics[width=\linewidth]{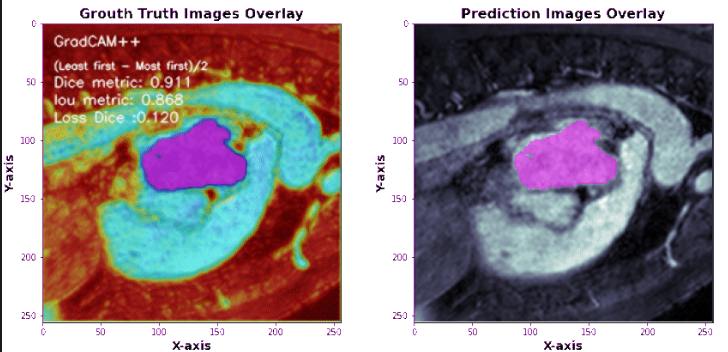}
  \vspace{0.5cm}
  \caption{Attention Gate Prediction Overlap within Mask}
\end{subfigure}
\vspace{0.5cm}
\caption{GradCAM++ Visualization - Attention Filter Gate vs. Attention Gate}
\label{fig:gradcam}
\end{figure}

In summary, the Attention Gate model outperformed other models in Mean Dice and Mean IoU scores, making it the preferred choice for image segmentation tasks. The Attention Filter Gate model also demonstrated competitive performance, leveraging the frequency domain for spatial dependency analysis.



\section{Discussion}

we explored our research approach, study design, and results, aiming to address image processing and segmentation challenges. Our focus lies in developing and evaluating novel techniques within the Global Filter Network (GFNet) to enhance image segmentation accuracy and efficiency. The Attention Filter Gate, inspired by the Fast Fourier Transform (FFT) and learnable weights, emerges as a promising approach. Comparative experiments with traditional models, including U-Net and Attention Gate, reveal the competitive performance of GFNet. Attention Gate demonstrates superior segmentation accuracy and consistency, while U-Net maintains reliability. Metrics such as Mean Dice and Mean Intersection over Union (IoU) provide deeper insights into the models' performance. Findings have implications for medical imaging and computer vision, suggesting that leveraging frequency domain information and learnable weights improves segmentation accuracy. Future research involves refining the Attention Filter Gate, exploring applications in different domains, and investigating interpretability and explainability. Delimitations and limitations, including dataset specificity, focus on one architectural variation, and hardware constraints, are acknowledged. Despite these, our research contributes valuable insights, opening avenues for future exploration in image segmentation. The Attention Filter Gate within the GFNet architecture represents a significant advancement with practical implications for researchers and practitioners in image processing.

\section{Conclusion}

In conclusion, our research introduces the Attention Filter Gate as a novel mechanism within the Global Filter Network (GFNet) for image segmentation tasks. The study demonstrates its competitiveness compared to traditional models and highlights the importance of frequency domain information in image processing. Our findings open avenues for future research and applications in various domains, emphasizing the potential for improved image segmentation techniques.

The development and evaluation of the Attention Filter Gate within the GFNet architecture represent significant contributions to the field of image processing, providing novel insights and practical implications for researchers and practitioners alike.

\section{Disclosures}
The authors declare that they have no conflict of interest

\section{Acknowledgments}
We would like to thank our respectful research assistant Moath Alawaqla, for his distinguished role of data collection. 
\section{Funding}
This work is supported by Jordan University of Science and Technology, Irbid-Jordan,


\bibliography{report} 
\vspace{2ex}
\bibliographystyle{spiejour}   

\end{document}